\documentclass[a4paper,11pt]{article}

\pdfoutput=1 

\usepackage{jcappub} 

\usepackage[T1]{fontenc} 

\usepackage{natbib} 

\title{Constraining the neutrino emission of gravitationally lensed Flat-Spectrum Radio Quasars with ANTARES data}

\emailAdd{manganos@ific.uv.es}
\emailAdd{elewyck@apc.univ-paris7.fr}

\author[a]{S.~Adri\'an-Mart\'inez}
\author[b]{A.~Albert}
\author[c]{M.~Andr\'e}
\author[d]{G.~Anton}
\author[a]{M.~Ardid}
\author[e]{J.-J.~Aubert}
\author[f]{B.~Baret}
\author[g]{J.~Barrios-Mart\'{\i}}
\author[h]{S.~Basa}
\author[e]{V.~Bertin}
% ...

\affiliation[a]{\scriptsize{Institut d'Investigaci\'o per a la Gesti\'o Integrada de les Zones Costaneres (IGIC) - Universitat Polit\`ecnica de Val\`encia. C/  Paranimf 1 , 46730 Gandia, Spain.}}
\affiliation[b]{\scriptsize{GRPHE - Institut universitaire de technologie de Colmar, 34 rue du Grillenbreit BP 50568 - 68008 Colmar, France}}
\affiliation[c]{\scriptsize{Technical University of Catalonia, Laboratory of Applied Bioacoustics, Rambla Exposici\'o,08800 Vilanova i la Geltr\'u,Barcelona, Spain}}
\affiliation[d]{\scriptsize{Friedrich-Alexander-Universit\"at Erlangen-N\"urnberg, Erlangen Centre for Astroparticle Physics, Erwin-Rommel-Str. 1, 91058 Erlangen, Germany}}
\affiliation[e]{\scriptsize{CPPM, Aix-Marseille Universit\'e, CNRS/IN2P3, Marseille, France}}
\affiliation[f]{\scriptsize{APC, Universit\'e Paris Diderot, CNRS/IN2P3, CEA/IRFU, Observatoire de Paris, Sorbonne Paris Cit\'e, 75205 Paris, France}}
\affiliation[g]{\scriptsize{IFIC - Instituto de F\'isica Corpuscular, Edificios Investigaci\'on de Paterna, CSIC - Universitat de Val\`encia, Apdo. de Correos 22085, 46071 Valencia, Spain}}
\affiliation[h]{\scriptsize{LAM - Laboratoire d'Astrophysique de Marseille, P\^ole de l'\'Etoile Site de Ch\^ateau-Gombert, rue Fr\'ed\'eric Joliot-Curie 38,  13388 Marseille Cedex 13, France}}
% ...

\abstract 
{
This paper proposes to exploit gravitational lensing effects to improve the sensitivity of neutrino telescopes to the intrinsic neutrino emission of distant blazar populations. This strategy is illustrated with a search for cosmic neutrinos in the direction of four distant and gravitationally lensed Flat-Spectrum Radio Quasars. The magnification factor is estimated for each system assuming a singular isothermal profile for the lens. Based on data collected from 2007 to 2012 by the ANTARES neutrino telescope, the strongest constraint is obtained from the lensed quasar B0218+357, providing a limit on the total neutrino \mbox{luminosity} of this source of $1.08\times 10^{46}\,\mathrm{erg}\,\mathrm{s}^{-1}$. This limit is about one order of magnitude lower than those previously obtained in the ANTARES standard point source searches with non-lensed Flat-Spectrum Radio Quasars.}

\begin{document}
\maketitle
\flushbottom




\section{Introduction}

Active Galactic Nuclei (AGNs) are longstanding candidate sources for the highest-energy cosmic rays, the origin of which is still unknown. 
The blazar subclass is comprised of BL\,Lac objects and the 
more luminous Flat-Spectrum Radio Quasars (FSRQs), 
the relativistic jets of which are pointed at a small angle to the line of
sight~\citep{1995PASP..107..803U}.
The matter content of AGN jets 
is still heavily debated, as both leptonic and hadronic models can in
principle explain the broad band spectral energy distribution of
blazars~\citep{boettcher07}. As of now, for electromagnetic radiation there is no
simple observable that allows us to distinguish between both types of
models. The detection of neutrinos from such jets might therefore help solve this 
long-standing issue.

In hadronic models for AGN jets, neutrinos are
predicted to be emitted along with gamma-rays by processes involving the interaction of
accelerated hadrons with the surrounding medium and radiation fields, and the subsequent production and decay of pions
and kaons (see e.g.~\cite{gaisser95,learned00} and references therein).
 The neutrino spectrum should therefore be closely related to the hadron spectrum, usually assumed to be a power-law with spectral index $\gamma \simeq -2$ as suggested by the theory of diffuse shock acceleration~\citep{Bell:1978zc,Bell:1978fj}. The IceCube neutrino telescope has recently provided evidence for a cosmic component in the diffuse high-energy neutrino flux~\citep{Aartsen:2013jdh}, part of which could originate from a population of unresolved extragalactic sources, possibly blazars~\citep{Krauss:2014tna}. Yet, no individual neutrino source has so far been identified, and, given the low expected neutrino fluxes, the published upper limits do not allow different jet models to be distinguished. BLLacs are interesting targets to derive generic constraints on the jet emission mechanisms as they can be found at relatively small redshifts. For blazar populations which are typically distributed at much larger distances, such as the FSRQs, we argue that  gravitational lensing can help enhance the sensitivity to neutrino emission. 

Gravitational lensing of electromagnetic radiation from distant
astrophysical sources is a well-known prediction of Einstein's theory of 
general relativity~\citep{Einstein,Zwicky}. Since the first detection of
multiple images of a gravitationally lensed quasar, QSO~0957+561~\citep{1979Nature,1979ApJ}, hundreds of lens systems have been
discovered and studied, opening new perspectives both in astrophysics
and cosmology~(see \cite{Bartelmann:2010fz} for a recent review).

Due to their very low masses ($m_\nu$ do not exceed $\sim$1~eV \cite{Beringer:1900zz}), 
neutrinos are expected to undergo the same phenomenon of gravitational lensing as photons. Possible configurations for neutrino lensing by massive astrophysical objects have been theoretically studied in the literature, e.g.\cite{Gerver,Elewyck2,Mena,Eiroa}. 
For distant neutrino sources gravitationally lensed by an interposed galaxy (or galaxy cluster),  
the magnification of the neutrino flux is expected to be equal to that of the photon flux.

Here, we make use of this similarity between the
lensing of photons and of neutrinos to significantly lower the upper
limits on neutrino emission from FSRQs, by using the luminosity boost provided by gravitational lenses that magnify some distant radio-loud AGN.
We illustrate this method by performing a search for an excess of neutrino events
in the direction of distant lensed FSRQs in the field of view of the ANTARES neutrino telescope. 

The remainder of this paper is organized as follows.
Section~\ref{sec:antares} describes the ANTARES neutrino telescope, which is 
used to detect neutrino events from gravitationally lensed sources, followed in Section~\ref{sec:addobjects} by a description of the candidate sample. We present
and discuss our results in Section~\ref{sec:results}. Specifically, we show that this strategy leads to an  improvement in the upper limit on the intrinsic neutrino luminosity of FSRQs. 

\section{The ANTARES Neutrino Telescope and Data Sample}\label{sec:antares}
The ANTARES neutrino telescope~\citep{Ant2011} is located in the Mediterranean Sea, about 40\,km offshore from Toulon, France ($42^\circ 48'$N, $6^\circ 10'$E).
The detector consists of 885 photomultiplier tubes (PMTs) mounted on
twelve vertical lines anchored to the seabed at a depth of 2475m, with a typical inter-line separation of $\sim$65\,m. 
The operation principle is based on the detection of
Cherenkov light induced by relativistic muons produced in the
charged-current interactions of high-energy neutrinos with matter
surrounding the detector. The knowledge of the time and amplitude of
the light pulses recorded by the PMTs allows the reconstruction of the
trajectory of the muon and to infer the arrival direction of the
incident neutrino. To limit the background from down-going
atmospheric muons, the design of ANTARES is optimised for the detection
of up-going muons produced by neutrinos which have traversed the
Earth. Its instantaneous field of view is $\sim 2\pi\,\mathrm{sr}$ for
neutrino energies $E_\nu \gtrsim
100\,\mathrm{GeV}$. Further detail on the above can be found in refs.~\cite{Ant2011,Ant2002,DAQ}.

The ANTARES collaboration has developed several strategies to search
for cosmic neutrino candidates, also in association with other
cosmic messengers such as cosmic-, X-, and gamma-rays, and gravitational waves (e.g.~\cite{Aguilar:2010ab,Ageron:2011pe,2012APh....36..204A,AdrianMartinez:2012tf,Adrian-Martinez:2013sga}).
A search for neutrino point-like sources was conducted with the data sample corresponding to the first four years of operation of the detector, 2007--2010~\cite{AdrianMartinez:2012rp}. This search has recently been extended to the years 2011--2012, for a total detector livetime of 1338 days~\cite{Adrian-Martinez:2014wzf}. The selection criteria 
have been optimised to search for $E^{-2}$ neutrino fluxes from point-like
astrophysical sources. Upgoing events have been selected to reject the bulk of background atmospheric muons. Additional cuts on
the quality of the muon track reconstruction have also been applied to eliminate events that correspond to downgoing atmospheric
muons which are misreconstructed as upgoing. Most of the remaining
events are atmospheric muon neutrinos which constitute the  
primary background for cosmic neutrino searches. 
The final  
sample contains 5516 neutrino candidates with a predicted atmospheric
muon neutrino purity of around 90\% and an estimated median angular
resolution of $0.38^\circ$~\citep{Adrian-Martinez:2014wzf}. 

The above-mentioned sample was searched for an excess neutrino flux in the direction of 51 candidate neutrino sources (including the Galactic Centre). This list of sources was established mainly on the basis of the
intensity of their gamma-ray emission as observed by \textsl{Fermi}~\citep{Atwood:2009ez} and H.E.S.S.~\citep{Bernloehr:2003vd} and 
includes five non-lensed FSRQs. No
statistically significant excess has been found in the direction of any of the candidate sources. The corresponding 90\% confidence level (C.L.)  upper limits on the neutrino flux, assuming an $E^{-2}$ spectrum, have been derived;  they are the most restrictive to date in a significant fraction of the Southern Sky. In the next section, we show how these limits can be improved upon by using lensed sources.


\section{Search for Neutrino Emission from Distant, Lensed Blazars}
\label{sec:addobjects}

\subsection{Determination of the Magnification Factor}
\label{sec:mu}
The strategy presented here relies on an estimation of the magnification factor for each lensed system, as obtained from photon observations. To estimate the magnification of a given lensed image, a model is required for the mass distribution of the lens.  Here, we utilise the simplest model able to account for the image morphology and
magnification ratios in each lens system: the singular
isothermal ellipsoid (SIE). SIE models are also a reasonable
description of the mass distributions of elliptical galaxies (e.g., \cite{Sonnenfeld:2013}). We
adjust the parameters of SIEs with the "gravlens" modeling software~\cite{Keeton} to fit image positions and flux ratios. We use image positions from infrared H-band HST
observations from the CASTLES project\footnote{www.cfa.harvard.edu/castles/},
which have uncertainties of $0.003$ arcsec. We assume the centre of
mass of each lens galaxy to be its centre of brightness. The centre for
each SIE model is then the centre of brightness of each lens
galaxy. The positions of the SIEs have uncertainties between
$0.01$ and $0.05$ arcsec (CASTLES). Because the flux ratios may be
influenced by microlensing, the SIEs cannot reproduce them to better
than a 10\% uncertainty, which we adopt for our flux
uncertainties. The resulting uncertainties for the estimates of
magnifications for the lensed images in our sample are about 15\% (10\% for quadruple-image systems),
excluding systematic uncertainties.

\subsection{Candidate Distant Lensed Blazars}
\label{sec:cand}
The most interesting lensed blazar candidates are B0218$+$357~\citep{1972AJ.....77..797P}  and  PKS 1830$-$211~\citep{1988MNRAS}: they 
are visible to ANTARES and have also been detected by \textsl{Fermi}. 
B0218$+$357 is a double-image lens system with 
an extended jet at redshift 
$z=0.96$, while the intervening lensing spiral galaxy has $z=0.68$~\citep{1992AJ....104.1320O}. We obtain magnification values of  
12.3 and 8.5 for the two images based on the SIE model. These estimates are consistent with the observed flux ratios at radio
wavelengths~\citep{Biggs:1998ez}. B0218$+$357 is also the first lensed system for which a clear $\gamma$-ray
measurement of a time delay for two images has been 
reported~\citep{Cheung:2014dma}. 
The measurement utilised flares detected
with the \textsl{Fermi} Large Area Telescope (LAT). Although the LAT is unable
to resolve the two images, the flares were of sufficient amplitude
(peaking at $\sim$ 20$-$50 times the quiescent flux) to conduct an
autocorrelation analysis. The result was a delay of $11.46\pm 0.16$
days, generally consistent with previous measurements
at radio wavelengths. 
Some of the uneven structures of the flares
implied that microlensing may be occurring in this system. To
establish the effect of microlensing on the magnifications of the
images that we use in this paper, models of the distribution of stars
in the lens galaxy and their kinematics are necessary.  Further analysis
is underway to model the behavior of the $\gamma$-ray flares (Falco et
al., in preparation). 

Flares in $\gamma$-rays from \textsl{Fermi}-LAT were also reported for
PKS~1830$-$211, a double-image lens system with a separation of 
\mbox{$\sim 1$~arcsec~\citep{Donna}}. This radio-loud blazar is at $z=2.51$ and is gravitationally lensed by a spiral galaxy at $z=0.89$. Based on the SIE model, the magnification values that we obtain for the two images are 3.7 and 1.5. A time delay of $27.1 \pm 0.6$ days between the images was estimated at radio wavelengths~\cite{Lovell:1998ka}. Evidence for a time delay in the \textsl{Fermi} LAT $\gamma$-ray data, consistent with the radio measurement, has also been reported for this source~\cite{2011A&A...528L...3B}. Observations with ALMA that overlapped with some of those of \textsl{Fermi} suggested microlensing at sub-mm wavelengths~\citep{Marti-Vidal:2013vva}.
 Their measurements also revealed chromatic variations in the flux densities of the images at
these wavelengths, which may arise from microlensing or from intrinsic
variability of the blazar jet. Our estimates for the magnifications of
the images of PKS~1830$-$211 yield a magnification ratio of $\sim 2.4$,
compared with the estimated magnification ratio of $\sim 1.5$ at radio
wavelengths~\cite{Lovell:1998ka}. Because this source is
highly variable (both extrinsically and intrinsically), the estimates
of magnification ratios are affected and also variable. We adopt
our model magnifications as fiducial values.

We also include in our study B1422$+$231~\citep{Raychaudhury:2003cf} and B1030$+$074~\citep{Xanthopoulos:1998zm} which are two lensed blazars in the ANTARES field of view, although with no associated \textsl{Fermi} detection so far. 
B1422$+$231 is a four-image quasar at  $z=3.62$
lensed by a galaxy at $z=0.34$. Its optical spectrum shows broad emission lines, and it is categorized as an FSRQ in the Multifrequency Catalogue of Blazars~\citep{Massaro:2010si}. The magnification values of the four images as obtained from the SIE model are 16.0, 15.0, 11.1 and 0.9. B1030$+$074 is a two-image system with a variable source exhibiting jet structure at  $z=1.54$, and a 
lens galaxy at $z=0.60$. It is a blazar of uncertain type in~\cite{Massaro:2010si}. We obtain magnification values of 2.4 and 0.27 for the two images.

\begin{table}
\centering
\begin{tabular}{lrrrrrrr}\\
\hline \\
Name 
                              & $z$ & $d_L$ & $\phi_{\nu}^{90CL}$ & $\mu$ (n) & $L^{90}$ \\\\
\hline 
\hline \\ 

3C~279           
                                          & 0.54 & 3.12 & 3.45 & 1 & 6.44\\
PKS~1454$-$354    
                                              &1.40  &10.2  & 1.70 & 1 & 33.8\\
3C~454.3          
                                                 &0.86  & 5.54 & 2.39 & 1 &14.1\\

PKS~1502$+$106    
                                                 &1.84  &14.3  & 2.31 & 1 &90.5\\

PKS~0727$-$11      
                                              &1.60  & 12.0 & 2.01 & 1 &55.3\\
\hline
PKS~1830$-$211    
                                  & 2.51 &  21.0 & 1.89  &   5.20 (2) &30.8\\
B0218$+$357  
                                  & 0.96 & 6.35  & 2.91 &  20.8 (2) & 1.08\\
B1422$+$231  
                                  & 3.62 &  32.7   &  2.46 & 43.0 (4)&11.7\\
B1030$+$074  
                                  &1.54  & 11.5 &  2.26 & 2.67 (2)& 21.5 \\
\hline
QSO~0957$+$561
                                  &1.41  & 10.3 &  - & 2.80 (2) & - \\
\hline \\
\end{tabular}
  \\

  \caption{Redshift $z$, luminosity distance $d_L$ (in units Gpc), 90\% C.L. flux limit (in units
 $10^{-8} \mathrm{GeV}\,
  \mathrm{cm}^{-2}\, \mathrm{s}^{-1}$),
 magnification factor $\mu$ and number of lensed images $n$ (when relevant), and upper limit on intrinsic luminosity $L^{90}$ (in units $10^{46}\,\mathrm{erg}\, \mathrm{s}^{-1}$) 
 for the FSRQs considered in this analysis. }
\label{tab:fsrqs}
\end{table}

\subsection{Search for Neutrino Emission}
\label{sec:search}
We search for neutrino emission from the above sources following the
same procedure as described in \cite{AdrianMartinez:2012rp,Adrian-Martinez:2014wzf}. Neutrino events within 20$^\circ$ of the source are used to construct an unbinned likelihood function including both signal and atmospheric background components. For each source, this likelihood is then maximised with respect to the number of signal events. The sensitivity of the analysis is evaluated through the generation of large numbers of pseudo-experiments simulating background and signal. No significant excess of neutrinos above the expected background has been found for the four candidate sources. 
The corresponding 90\% C.L. upper limits on the neutrino flux have been derived using the Feldman \& Cousins approach~\cite{Feldman:1997qc}; they are given in Table~\ref{tab:fsrqs} along with the limits  on the five non-lensed FSRQs obtained in the previous ANTARES analysis. 

\begin{figure}\centering
  \includegraphics[width=\hsize]{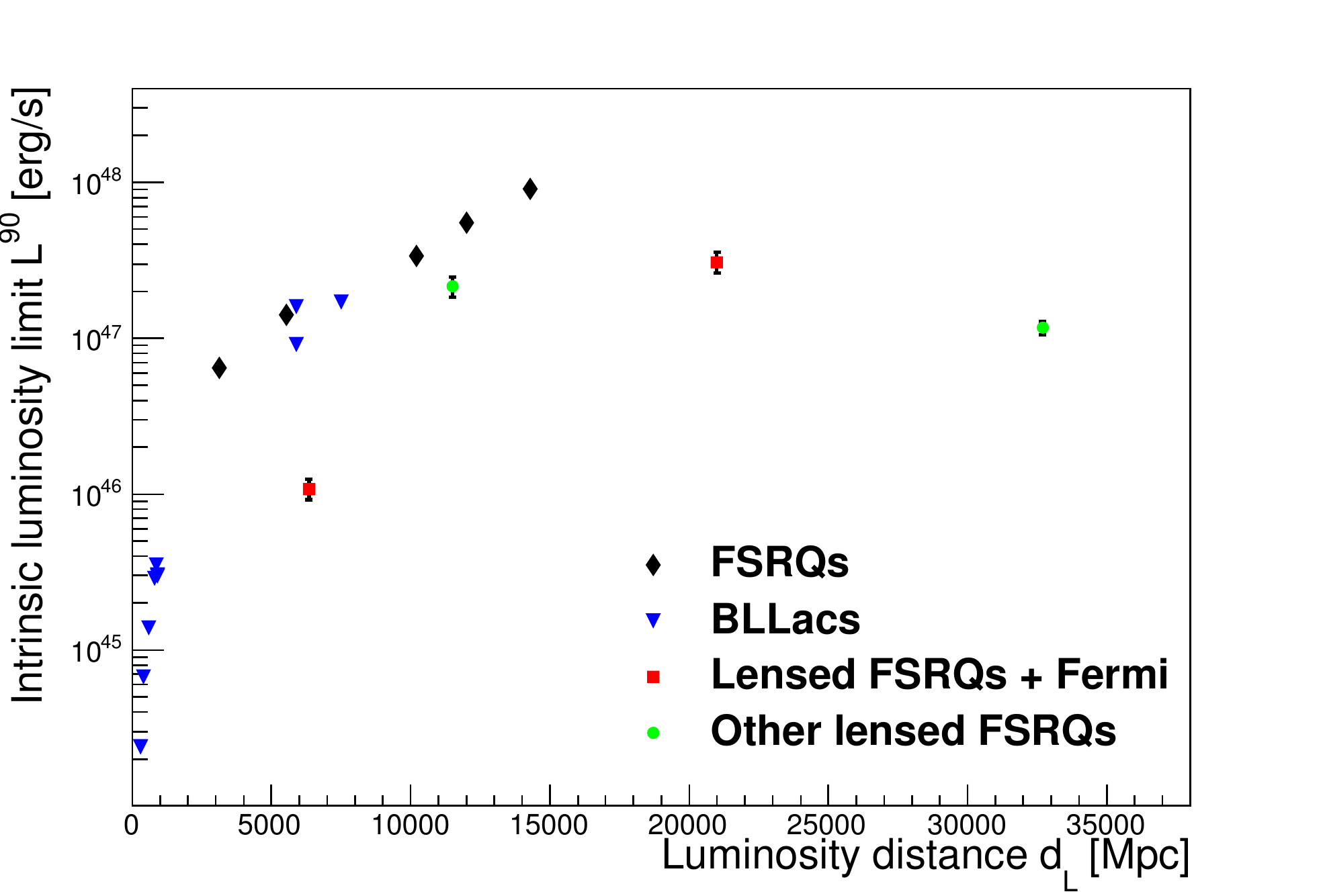}
  \caption{Limit on the instrinsic neutrino luminosity $L^{90}$ of blazars viewed by ANTARES as a function of their luminosity distance. The black diamonds (resp. blue triangles)  correspond to non-lensed FSRQs (resp. BLLacs) in the list of candidate sources of \cite{Adrian-Martinez:2014wzf}.  The red squares correspond to the two lensed FSRQs also seen by \textsl{Fermi} (B0218$+$357 and  PKS 1830$-$211) and the green circles correspond to the two other lensed blazars considered in this paper (B1422$+$231 and B1030$+$074). The error bars on the lensed sources represent the uncertainty in the determination of the magnification factor.}
  \label{fig}
\end{figure}

 \section{Results and Discussion}\label{sec:results}
 
 From the upper limits on the neutrino flux at Earth $\phi_{\nu}^{90CL}$ (given in Table~\ref{tab:fsrqs}), we derive limits on the intrinsic neutrino luminosities of the sources. The total isotropic emitted power in high-energy neutrinos ($\gtrsim 1$\,TeV) for a blazar at luminosity distance $d_L$ is
 \begin{equation}
 L^{90} = \frac{1}{\mu}\, 4\pi d_L^2\, \phi_{\nu}^{90CL},
 \end{equation}
 where $\mu$ is the magnification factor ($\mu = 1$ for non-lensed sources). The corresponding values are given in Table~\ref{tab:fsrqs} for both lensed and non-lensed FSRQs considered in this study.  
Note that the value $L^{90}$ itself is not a source-intrinsic
quantity. It depends on the viewing angle via the Doppler factor. In lack of
well-constrained viewing angles for most blazars, however, conclusions about
the intrinsic luminosities can be derived by applying statistical methods~\citep{Cohen2007}.

The results are summarised in Figure~\ref{fig} where the intrinsic luminosity limit is plotted against the luminosity distance of the source. For the lensed systems, the error bar account for the uncertainty in the determination of the magnification factor as discussed in Sect.~\ref{sec:mu}. We have added for comparison the luminosity limits obtained for the BLLacs in the ANTARES list of sources, most of which are located at a much closer distance~\citep{Adrian-Martinez:2014wzf}. The isotropic power limits obtained for the non-lensed FSRQs are between $6 \times 10^{46}$ and $9 \times 10^{47}\,\mathrm{erg}\,\mathrm{s}^{-1}$.   
 These powers can be compared with the bolometric luminosities of AGNs, which are typically in the range $10^{44}$--$10^{47}\,\mathrm{erg}\,\mathrm{s}^{-1}$~\citep{Woo:2002un,MNR:MNR21513,boettcher07}, but can rise up to $10^{49}\,\mathrm{erg}\,\mathrm{s}^{-1}$ for some hadronic jet models such as in the synchrotron proton blazar interpretation of 3C~279~\cite{Boettcher:2008tq}. 

One directly sees from Figure~\ref{fig} that the limits derived from lensed FSRQs are stronger than those corresponding to non-lensed sources of the same class at comparable distances, and that lensing can be efficiently used to improve the constraints on 
neutrino emission from FSRQs. The strongest limit is obtained for the \textsl{Fermi}-detected B0218$+$357; it corresponds to a total neutrino power of $1.1\times 10^{46}\,\mathrm{erg}\,\mathrm{s}^{-1}$, about one order of magnitude lower than the lowest limit achieved with non-lensed FSRQs. 
This limit is expected to improve in the future, in particular when the multi-km$^3$ neutrino telescope KM3NeT~\citep{KM3NeTTDR}, with an instrumented volume about 100 times bigger than that of ANTARES, becomes operational in the Mediterranean.

A similar study could in fact be performed with the lensed quasar QSO~0957$+$561, a doubly-imaged, wide-separation system with the source at $z = 1.41$ and the lensing galaxy at $z=0.36$~\citep{1979Nature,1979ApJ}. 
This quasar is not in the ANTARES field of view, but its neutrino emission could be constrained by IceCube; the information for this source is therefore also 
included in Table~\ref{tab:fsrqs}. 
Assuming
a typical ANTARES sensitivity for a similar source in the ANTARES field of
view, as given in~\cite{Adrian-Martinez:2014wzf}, the 90\% C.L. upper limit on the neutrino luminosity for QSO~0957$+$561  
would be $L^{90}\simeq 1.8\times 10^{47}\,\mathrm{erg}\,\mathrm{s}^{-1}$.
Based on the sensitivities presented in~\cite{Abbasi:2010rd}, 
the limit set by IceCube would approximately be 10 times stronger, 
and thus comparable to the 
limit obtained by ANTARES on B0218$+$357. 

In conclusion, we suggest that neutrino telescopes include the lensed FSRQs discussed above in their future searches for steady point-source neutrino emission.

\acknowledgments

The authors would like to thank D. Allard, O. Mena, J. A. Mu\~noz and 
G. E. Romero for enlightening discussions during the preparation of this manuscript.  

They also acknowledge the financial support of the funding agencies:
Centre National de la Recherche Scientifique (CNRS),
Commissariat \`a l'\'energie atomique et aux \'energies alternatives (CEA),
Commission Europ\'eenne (FEDER fund and Marie Curie Program),
R\'egion Alsace (contrat CPER), R\'egion
Provence-Alpes-C\^ote d'Azur, D\'e\-par\-tement du Var and Ville de La
Seyne-sur-Mer, France;
Bundesministerium f\"ur Bildung und Forschung (BMBF), Germany;
Istituto Nazionale di Fisica Nucleare (INFN), Italy;
Stichting voor Fundamenteel Onderzoek der Materie (FOM), Nederlandse
organisatie voor Wetenschappelijk Onderzoek (NWO), the Netherlands;
Council of the President of the Russian Federation for young
scientists and leading scientific schools supporting grants, Russia;
National Authority for Scientific Research (UEFISCDI), Romania;
Servicio P\'ublico de Empleo Estatal (SEPE), Ministerio de Ciencia e Innovaci\'on (MICINN), 
Prometeo of Generalitat Valenciana and MultiDark, Spain;
Agence de l'Oriental and CNRST, Morocco.
We also acknowledge the technical support of Ifremer, AIM and Foselev Marine for 
the sea operation and the CC-IN2P3 for the computing facilities.


\bibliographystyle{JHEP}
\bibliography{./lensingbib_final}   

\end{document}